\documentclass[acmsmall,nonacm]{acmart}

\usepackage{enumitem}
\usepackage{parskip}
\usepackage{caption}
\usepackage{wrapfig}
\usepackage{comment}
\usepackage{multirow}
\usepackage{colortbl}

\newlist{questions}{enumerate}{2}
\setlist[questions,1]{label=RQ\arabic*.,ref=RQ\arabic*}

\renewcommand\footnotetextcopyrightpermission[1]{} 

\begin{document}

\title{Digital Contact Tracing: Examining the Effects of Understanding and Release Organization on Public Trust}
\subtitle{Undergraduate Honors Research Paper}

\author{Lucas Draper}
 \affiliation{%
   \institution{Oberlin College}
   \city{Oberlin, OH}
   \country{USA}}
 \email{lucas.draper@oberlin.edu}

\begin{abstract}

Contact tracing has existed in various forms for a very long time. With the rise of COVID-19, the concept has become increasingly important to help slow the spread of the virus. One approach to modernizing contact tracing is to introduce applications that detect all close contacts without individuals having to interact knowingly. 101 United States adults were surveyed in June of 2022 regarding their perceptions and trust of COVID-19 contact tracing applications. We see no definitive correlation between an individual’s understanding of privacy protection procedures for contact tracing applications and their willingness to trust such an application. We also see that the release of the application by a private entity like Google-Apple or by a public entity like the United States Federal Government has no significant correlation with a person’s trust in the application. 
\end{abstract}


\maketitle

\section{Introduction}
COVID-19 has changed how we think about our interactions. Before vaccines, every interaction between two people was a risk of infection, and with the virus's incubation period and asymptomatic infection being a possibility, COVID-19 spread rapidly. 

Various governments and private entities began attempting to build applications that utilize Bluetooth or GPS technology in a smartphone to maintain a list of close contacts, allowing faster and more effective contact tracing and, hopefully, a reduction in COVID-19 infections. One of the most prominent producers of contact tracing infrastructure was the collaboration of Google LLC. and Apple Inc. to create a cross-platform application programming interface (API) upon which various government agencies could construct contact tracing applications. The specifics of this API are highly guarded to prevent malicious agents from exploiting any flaws in the framework. 

The basics of the framework are laid out in layperson's terms in Appendix A, which is the statement given to our participants explaining how it operates. 

The API operates over Bluetooth and does not use the device's GPS location at any point. As outlined in the Cryptography specifications document released by Google-Apple, there are many layers to the application's contact identification system. Along with generating the exposure keys used to identify a person's contact, the application also generates an associated metadata key, the details of which are not elaborated on in the documentation and will be ignored in this explanation. 

Each device generates a new Temporary Exposure Key (TEK) daily and stores it along with the interval in which it was created. This interval is determined relative to the current Unix Epoch Time at the time of generation. The TEK is then used to generate a new Rolling Proximity Identifier (RPI) every time the Bluetooth MAC address changes (every 15-20 minutes) to prevent third parties from tracking a specific user. These RPIs are encrypted using a cryptography function with a 128-bit key, making it extremely difficult to decrypt. This RPI is broadcast to all devices near the user. When the device comes into contact with another device using the contact tracing application, RPIs are exchanged. Each device stores the last 14 days of TEKs and the interval at which they were created in case the user tests positive. 

Once a person tests positive, the last 14 TEKs are uploaded to a central server. Other devices frequently poll this central server, locally calculating all RPIs from the provided TEKs. If an RPI they have been exposed to appears on the central server, the associated metadata is accessed. If the exposure is significant enough to be a possible infection, the user is notified \cite{Cryptography}. 

The technology behind this API is complicated, and it is unreasonable to expect that an average user would understand the intricacies of its privacy protection methods. Because of this, it is reasonable to assume Apple and Google have access to the identifying information of a user that could be utilized for malicious purposes. This assumption is what led to the development of our research questions.

\begin{questions}[itemindent=1em] 
    \item \textit{Does having an understanding of the privacy preservation methods of the Google-Apple Exposure Notification (GAEN) API affect users' trust?}
    \item \textit{Does the involvement of private companies influence the users’ trust in their privacy protection?}
\end{questions}

We hypothesized that RQ1 would see an increase in trust in those who received the explanation compared to those who did not. We suspect a critical factor in the general mistrust of this technology is a lack of understanding regarding the privacy protection methods used. There are also many false assumptions within the general public, such as these applications using GPS location to track contact, which causes many people to distrust this technology. 

Regarding RQ2, we suspect that users have a higher distrust of the government and, as a result, will trust a privately developed application more than an application created by a government agency. This distrust stems from various sources but was furthered by situations such as the NSA's mass surveillance of the American public. 

Although the COVID-19 pandemic has become an accepted part of 'normal' life, investigating the acceptance of contact tracing technologies is a valuable pursuit, as there will likely be another pandemic at some point in the future. This information can be used to develop a more widely accepted application for preventing infection during this new pandemic.  

\section{Related Work}
In this section, we will discuss the plethora of research that predates this paper regarding digital contact tracing and its interaction with the digital privacy sphere.

One of the critical areas of computer science research regarding COVID-19 was the development of contact tracing applications. Much of this research was dedicated to understanding public perceptions of contact tracing technologies and what affects users' decisions to download these apps. It has been suggested but not confirmed from prior research that user understanding of privacy protection infrastructure positively affects their willingness to use such applications, which is one of our two research questions. There are also suggestions that release organization affects public trust, but this has not been investigated in great detail. This knowledge gap is what led to the determination of our research questions. 

There are a variety of factors that influence people's willingness to download contact tracing applications. A study by Altmann et al. found that participants with a higher distrust of the government also had higher rates of concern about government surveillance through such technology \cite{Altmann}. This correlation makes sense, as these individuals view the government as an enemy and, as a result, would expect their information to be used for nefarious purposes. Altmann also found that using Bluetooth technology or GPS location to determine contact had little effect on the public's willingness to use the application. They also report that an opt-out model would have a higher uptake than an opt-in application. However, we question this model's efficacy, as most users may need to be made aware of the presence of the service or how to opt out. 

There is a disagreement in the research as to whether the public supports this kind of contact tracing application. Altmann found widespread support for the technology across all demographics they surveyed. However, Wang and Bashir found only 63\% of their participants felt it was "acceptable if the government is tracking the location of COVID-19 cases" \cite{WangBashir}. Similarly, Maytin et al. found that, in their survey of young adults, only 45\% would share their COVID-19 status, and 33\% would be willing to allow access to their cell phones for passive tracking \cite{Maytin}.  

Prior research has also discovered that salient COVID-19 concerns reduce the willingness to adopt such technologies. Chan and Saqib surveyed three different populations, first in France, then in Australia, and finally in the US, to determine the effects of COVID-19 prevalence on public trust. In all three cases, when concerns regarding COVID-19 were high, there was an overall decrease in willingness to use tracing applications. They suggest this is due to "greater social conservatism, presumably including greater emphasis on personal privacy, thereby trumping privacy concerns" \cite{ChanSaqib}. Calloway et al. had similar findings. However, they note salient COVID-19 concerns "did not shift [the] perceived privacy risks" of participants \cite{Calloway}. 

Research has also determined that the public would rather trust a government organization with its contact tracing information over a private entity like Apple or Microsoft. Maytin et al. found that young respondents "showed overwhelming distrust" towards private companies, and only 15.2\% would share their information with these companies for contact tracing purposes \cite{Maytin}. This was supported by Calloway et al., who found their participants were more comfortable sharing their information with law enforcement than with private companies for contact tracing \cite{Calloway}. 

Kaptchuk et al. found that a factor affecting the general willingness to download was the accuracy of the application. Overall they found respondents were "8\% less likely to install an app with false negatives", and the risk of false positives or false negatives (31\% and 32\%, respectively) ranked higher in respondent concern than a privacy leak (17\%) \cite{Kaptchuk}. 

There are also a variety of demographic areas that influence a person's willingness to utilize an application designed for contact tracing. Kaptchuk et al. found that those who knew someone who died from COVID-19 were over five times as likely to be willing to install an application with privacy risks. They also found that those who self-identify as Democrats were nearly three times as likely to be willing to install an app than those who identified as Republicans \cite{Kaptchuk}. This statistic was supported by Grande et al. \cite{Grande} and Zhang et al. \cite{Zhang}. However, Grande found somewhat contradictory results for Kaptchuk's first observation, noting that personal experience with COVID-19, either personal infection or infection of a family member) were not generally associated with support \cite{Grande}. Li et al. also found that high household income and higher education levels had notable positive effects on intentions to install contact tracing applications. However, they did see that older people had notably lower intentions to install the application \cite{Li}. 

Research on the Google-Apple Exposure Notification framework has also produced some interesting results. Grande et al. found that 40\% of their respondents were willing to use an application developed under this framework, and sharing data with public health departments saw no decrease in support. They also saw a decrease in support when posing mandatory use of the theoretical application \cite{Grande}. 

Although not specific to the GAEN framework, both Zhang et al. and Li et al. investigated the effects of a decentralized data architecture (the kind GAEN uses) compared with a centralized data architecture and found conflicting results. Zhang et al. found that acceptance of a decentralized architecture was "5.4 percentage points higher on average" than a centralized architecture and that 44\% of respondents would download an application that stored data on users' phones, versus only 39\% who would install the application if it stored data on a centralized server \cite{Zhang}. However, Li et al. found that "app design choices such as decentralized vs. centralized architecture, location use, who provides the app, and disclosures about app security risks had very small effects on participants' adoption intentions" but noted that "a decentralized architecture could help reduce security risks that people are more concerned about" \cite{Li}. 

One study by Seberger and Patil grouped participants into 'individualists' and 'collectivists.' Individualists are defined as those who "give priority to their personal goals over the goals of their communities and behave primarily on the basis of individual attitudes rather than community norms" \cite{Triandis}. In comparison, a collectivist is a person who is "interdependent within their [community], give priority to the goals of their [society], shape their behavior primarily on the basis of [social] norms, and behave in a communal way" \cite{Triandis}. Seberger and Patil found that those who they classified as collectivists had a higher perceived benefit from contact tracing applications than those they classified as individualists. Individualists could also not be swayed by the framing of a situation, standing firm on their initial beliefs regarding such applications \cite{Seberger}. 

Overall, the collective body of research has explored a vast majority of the intricacies of public opinion and what leads individuals to accept or deny access to their personal interactions and location for pandemic exposure control. However, there has been little research regarding whether the lack of knowledge from the layperson or the release organization affects their willingness to utilize such technologies. 

\section{Methods}
In this section, we will discuss the methods behind both surveys we conducted, one to establish the clarity of our explanation and our primary survey to establish the effect of understanding and releasing organization on public trust. 

\subsection{Comprehension of Explanation}
To begin examining the effect of understanding on the public's willingness to trust COVID-19 contact tracing applications, we began by establishing a sufficient explanation of the contact tracing application and its privacy protection methods. We began by running a preliminary survey of 30 participants, presenting them with an explanation (see Appendix A for the final iteration of our explanation) and five comprehension questions to establish the effectiveness of the explanation. 

The questions first asked participants to order seven steps of a hypothetical COVID-19 exposure interaction and how the application handles this interaction. This was pulled directly from the example they were provided in the prior explanation. They were then asked if the contact tracing application tracked GPS location, whether any information uploaded upon a positive case could be tied to a specific individual by a third party, whether a close contact could determine a person's identity without a positive case and if a person’s information is uploaded without them testing positive. The only question of the five that we discounted when scoring accuracy was the ordering question, as a majority of participants did not get all seven steps in the correct order, and we could not definitively conclude that this was due to a lack of understanding. We also asked participants if there were any areas of the explanation that needed more clarity and if they had any further comments. 

After we ran this survey, we found that only 21 of our 30 participants answered 3 or more of the comprehension questions correctly, with only 7 participants getting all 5 correct. When we discount the ordering question, we see that 12 of our respondents obtained 100\% accuracy. We reviewed the feedback from participants on what area of the explanation needed more clarification. We then adapted the explanation to be presented in a bulleted format in the hopes this would increase user understanding. This revised explanation was piloted with another 30 participants. We saw 20 participants answered 3 or more correctly, and 7 participants got all 5 correct. When discounting the ordering question, we saw 14 participants scored 100\%. 

We deemed our second iteration of the explanation to have been more successful for two reasons. The first was the higher scores when disregarding the ordering question. The second was an analysis of the qualitative feedback we received. In the first group, most of the comments surrounded the hypothetical scenario being confusing. In contrast, the second group of participants mostly commented on the application being abused, which did not help improve explanation clarity.

\subsection{Explanation's Effect on Trust}
To examine both the effect of understanding on user trust, as well as the effect of release organization, we decided to split our participants into four groups. We surveyed a total of 101 people. The table below demonstrates the groups, how many participants they had, and what information they received:  
\begin{align*}
    \begin{tabular}{|c|c|c|}
    \hline
         & Google-Apple & Government \\
         \hline
        Explanation & Group 1 (23) & Group 3 (23) \\
        \hline
        No Explanation & Group 2 (28) & Group 4 (27) \\
        \hline
    \end{tabular}
\end{align*}

We primed half our population (groups 1 and 3) with the explanation we prepared in the earlier survey, swapping out the release organization for half of this smaller population to say ‘the United States Federal Government’ instead of Google-Apple (group 3). Groups 1 and 3 also received the same comprehension questions as the prior survey to establish baseline understanding allowing us to assess its impact correctly. Groups 2 and 4 received a vague statement stating the existence of a COVID-19 tracker developed by either Google-Apple or the US Federal Government and that it is used worldwide. 

All four groups were then asked if they trusted the application they were presented with and if they would trust the alternative more. They were also asked if they would download the application in three scenarios: voluntarily, if their local government required it and if endorsed by a celebrity they admired. 

We then asked the 16 questions laid out by Egelman et al. in their paper Scaling the Security Wall \cite{Egelman}. These questions were designed to gain insight into how privacy-conscious a person is and what steps they take to protect their privacy on the web. They can then be averaged to rank participants from most privacy-conscious to least privacy-conscious.  

\section{Results}
This section will present our study's results, demonstrating notable tallies and statistical data analysis. In the tables accompanying this section, please note 'Explain' refers to those who received the explanation, 'None' refers to those who did not receive the explanation, 'GAEN' refers to those who were told Google-Apple released the application, and 'Gov' refers to those who were told the Federal Government released the application. 

\subsection{Demographics}
\begin{table}[ht]
    \begin{tabular}{c c}
            \hline
            \textbf{Metric} & \textbf{Total} \\
            \hline
            Male & 42 \\
            Female & 55 \\
            Non-Binary & 4\\
            \hline
            18-25 & 23 \\
            26-35 & 46 \\
            36-45 & 17\\
            46-55 & 9\\
            55+ & 6\\
            \hline
            White & 87 \\
            African-American & 8 \\
            Asian & 4\\
            Pacific Islander & 1\\
            Other & 4\\
             & \\
        \end{tabular}     
    \hfill
        \begin{tabular}{c c}
            \hline
            \textbf{Metric} & \textbf{Total} \\
            \hline
            Democrat & 51 \\
            Independent & 26 \\
            Republican & 16 \\
            Libertarian & 1 \\
            No Preference & 7 \\
            \hline
            Associate degree & 11 \\
            Bachelor degree & 43 \\
            Graduate degree & 15 \\
            High school diploma & 11 \\
            Some college & 21 \\
            \hline
            Vaccinated, no booster & 31 \\
            Vaccinated and boosted & 46 \\
            Not vaccinated & 21 \\
            Declined to answer & 2 \\
        \end{tabular}
        \hfill
        \begin{tabular}{c c}
            \hline
            \textbf{Metric} & \textbf{Total} \\
            \hline
            Associate infected & 89 \\
            No associate infected & 11 \\
            Declined to answer & 1 \\
            \hline
            State application & 27 \\
            No state application & 16 \\
            Unsure & 58 \\
            \hline
            Installed, but deleted & 10 \\
            Installed and retained & 8 \\
            Did not install & 9 \\
             & \\
             & \\
             & \\
             & \\
             & \\
        \end{tabular}
        \caption{Demographics}
\end{table}

From our sample of 101 participants, 55 identified as female, 42 as male, and 4 as non-binary or third gender. 26-35 was the most prominent age range, with 46 participants, followed by 18-25 with 23, 36-45 with 17, and 46+ had a combined total of 15. 87 of our participants identified as white, 8 as African American, and 4 as Asian. 43 of our participants achieved a bachelor's degree as their highest level of education, 15 had a graduate's degree, 32 had either an associate's degree or started but did not complete college, and 11 only had a high school diploma. 51 of our participants identified as democrats, 26 as independents, and 16 identified as republicans. When asked about their vaccination status, 77 said they were vaccinated, with 46 of those receiving the booster vaccination. 21 said they were not vaccinated, and 2 declined to answer. 89 respondents said they knew someone who had been infected with COVID-19. Most of our respondents said they were unsure if their state had a contact tracing application (58 participants), while 27 responded in the affirmative and 16 were negative. Of the 27 that said yes, 18 downloaded the application, and 8 still had the application at the time of completing the survey. 

\newpage
\subsection{Situational Trust Scenarios}
\begin{wraptable}{l}{5cm}
    \vspace{-0.4cm}
    \centering
    \begin{tabular}{|c|c|c|c|}
    \hline
         & GAEN & Gov & Total \\
         \hline
        Explain & 11 & 10 & 21 \\
        \hline
        None & 10 & 15 & 25 \\
        \hline
        Total & 21 & 25 & 46 \\
        \hline
    \end{tabular}
    \caption*{Voluntary Download}
    \vspace{-0.7cm}
\end{wraptable}

46 of our total 101 participants (46\%) said they would download the application voluntarily. 21 of these received the explanation, and 25 did not. When separating by release organization, we see that 21 (41\%) of our Google-Apple participants would download the application voluntarily, while 25 (50\%) of the Federal Government participants would do the same.

\begin{wraptable}{r}{5cm}
    \centering
    \begin{tabular}{|c|c|c|c|}
    \hline
         & GAEN & Gov & Total \\
         \hline
        Explain & 14 & 14 & 28 \\
        \hline
        None & 20 & 19 & 39 \\
        \hline
        Total & 34 & 33 & 67 \\
        \hline
    \end{tabular}
    \caption*{Required By Local Government}
    \vspace{-0.7cm}
\end{wraptable}

67 of our participants (66\%) said they would download the application if required by their local government. Of these 67 participants, 28 received the explanation, and 39 did not. When comparing release organizations, we see 34 (67\%) of Google-Apple participants would abide by required downloads, and 33 (66\%) of Federal Government participants would download the application if their local government required them to do so.  

\begin{wraptable}{l}{5cm}
    \centering
    \begin{tabular}{|c|c|c|c|}
    \hline
         & GAEN & Gov & Total \\
         \hline
        Explain & 4 & 1 & 5 \\
        \hline
        None & 1 & 8 & 9 \\
        \hline
        Total & 5 & 9 & 14 \\
        \hline
    \end{tabular}
    \caption*{Endorsed By Celebrity}
    \vspace{-0.7cm}
\end{wraptable}

Lastly, participants were asked if they would download the application if it were endorsed by a celebrity they admired. Overall, 14 of our 101 participants said they would download it if endorsed by a celebrity, 5 of these received the explanation, and 9 did not. 5 of our Google-Apple respondents (10\%) stated they would use the application if a celebrity promoted it, as did 9 of our government respondents (18\%).

\subsection{RQ1: Does Understanding Affect User Trust}
Firstly, we will present the results relevant to our first research question, starting with examining the understanding of our participants.

\begin{wraptable}{l}{5.5cm}
    \vspace{-0.4cm}
    \centering
    \begin{tabular}{|c|c|c|c|}
    \hline
        Correct & Before & After & Improve \\
         \hline
        0 & 2.8 & 4 & 1.2 \\
        \hline
        1 & 3 & 3.75 & 0.75 \\
        \hline
        2 & 2 & 3.2 & 1.2\\
        \hline
        3 & 3 & 3.9 & 0.9 \\
        \hline
        4 & 3.2 & 4.4 & 1.1 \\
        \hline
    \end{tabular}
    \caption{Average Self-Reported Understanding Scores by Number of Correct Comprehension Questions}
    \vspace{-0.7cm}
\end{wraptable}

Of the 46 participants who received the explanation, only 29 (63\%) scored above 50\% on the understanding check questions. Only 10 (22\%) scored over 80\%. Of this subsection, 39 participants (85\%) self-reported that they either possibly or definitely understood how the application protected their privacy after receiving the explanation, although 23 (59\%) of these participants stated they understood the application before receiving the explanation. Overall, the explanation saw an average of 1.065 degrees of improvement on the standard 5-point Likert scale, with 25 (54\%) participants seeing an increase in their self-reported understanding. The average scores of each correct answer for the comprehension questions can be seen in Table 2.  

We do not see a relationship when we compare a person’s understanding of the explanation with their willingness to trust their given application. This result was proven via a t-test, which returned a p-value of 0.4629 (t = 0.74254, df = 33.932).

We see that those who received the explanation of the privacy protection method overall had 16 responses stating they would trust the application they were presented with. 10 of those being Google-Apple and 6 being from the Government group. In comparison, we see that only 12 respondents said they would trust the alternative application more, 7 from Google-Apple, and 5 from the Government group. Only 4 respondents in the Google-Apple group said they would trust the Google-Apple application but would trust the Government application more, compared with 3 from the government group who had the inverse view. 

\begin{wrapfigure}{r}{0.5\textwidth}
  \begin{center}
    \includegraphics[width=0.48\textwidth]{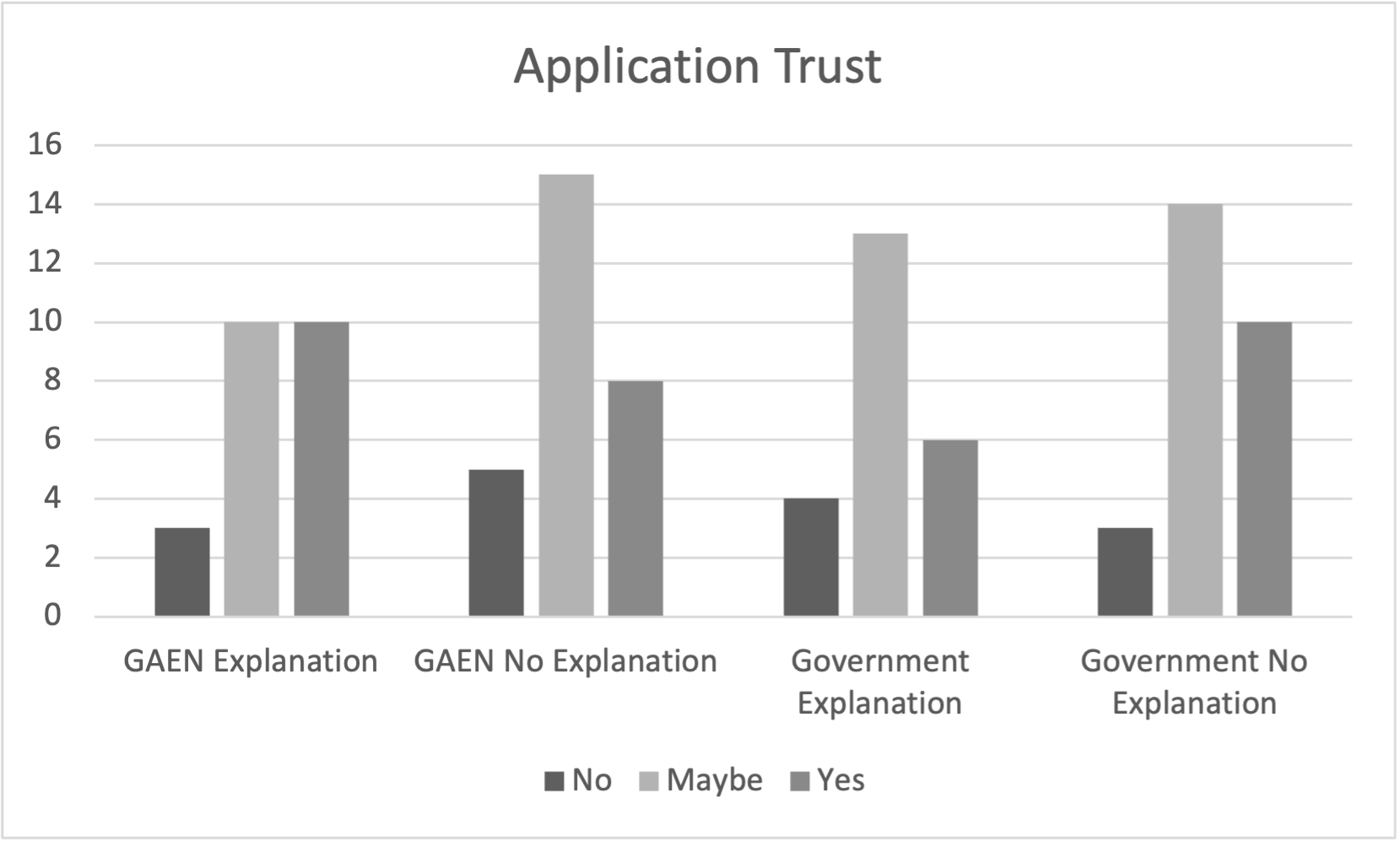}
  \end{center}
  \caption{Application Trust}
  \vspace{-0.7cm}
\end{wrapfigure}
When we analyze the results of those who did not receive the explanation, we see 18 who would trust the application they were presented with, 8 from the Google-Apple group, and 10 from the government group. We also see 20 people who would trust the other option more, 11 from the Google-Apple group and 9 from the government group. 4 members of both the government and Google-Apple groups members would trust their application but would trust the alternative more. These comparisons are presented in Figure 1.

\vspace{1cm}
We then investigated the effect of receiving the explanation on the users’ trust in their given organization. A t-test of these two groups returned a p-value of p-value = 0.919 (t = 0.10198, df = 94.937), and a Wilcoxon test returned 0.9042. The Chi-squared test returned a value of 0.9631 (X-squared = 0.075239, df = 2). For their trust of the alternative application, a t-test returned a p-value of 0.1337 (t = -1.5126, df = 95.75), the Wilcoxon test returned 0.1302, and the chi-squared test returned a 0.2894 (X-squared = 2.4798, df = 2).  

When we examine this separation within the two application groups, we see a similar lack of significant effect. For the Google-Apple application groups, we ran a t-test of their trust in the GAEN application when separated by whether they received the explanation and saw a p-value of 0.3185 (t = 1.0083, df = 46.611). A Wilcoxon test also saw inconclusive results with a p-value of 0.3002. A Chi-Squared test returned a warning that the p-value may not be reliable but returned a p-value of 0.5369 (X-squared = 1.244, df = 2). For the same group, when examining their trust in the government application, a t-test returned a p-value of 0.2476 (t = -1.1709, df = 46.354), while a Wilcoxon test returned 0.2316. The Chi-squared test for this data subset also gave the same warning but returned a p-value of 0.2992 (X-squared = 2.4132, df = 2). 

When separated by whether they received the explanation, the Government application groups saw a p-value of 0.3642 (t = -0.91639, df = 46.452) from a t-test, and the Wilcoxon test saw a p-value of 0.3629. This Chi-Squared test gave the same warning as the GAEN group, returning a p-value of 0.6487 (X-squared = 0.86543, df = 2). For the same group, when examining their trust in the GAEN application, a t-test returned a p-value of 0.3524 (t = -0.93912, df = 47.25), while a Wilcoxon test returned a p-value of 0.357. The Chi-squared test also gave a warning but returned a p-value of 0.6249 (X-squared = 0.94033, df = 2). 

\subsection{RQ2: Does Release Organization Affect User Trust?}
Next, we will present the results relevant to our second research question. We asked subjects if they would trust the application they were presented with and whether they would trust the alternative application more. 

\begin{table}[ht]
    \parbox{.30\linewidth}{
        \centering
        \begin{tabular}{|c|c|c|c|}
            \hline
            & Yes & Unsure & No \\
             \hline
            GAEN & 18 & 25 & 8 \\
            \hline
            Gov & 18 & 10 & 23 \\
            \hline
        \end{tabular}
        \caption*{GAEN Groups Trust}
    }
    \hfill
    \parbox{.30\linewidth}{
        \centering
        \begin{tabular}{|c|c|c|c|}
        \hline
             & Yes & Unsure & No \\
             \hline
            GAEN & 10 & 10 & 3 \\
            \hline
            Gov & 7 & 3 & 13 \\
            \hline
        \end{tabular}
        \caption*{GAEN Explanation Group Trust}
    }
    \hfill
    \parbox{.30\linewidth}{
        \centering
        \begin{tabular}{|c|c|c|c|}
            \hline
            & Yes & Unsure & No \\
            \hline
            GAEN & 8 & 15 & 5 \\
            \hline
            Gov & 11 & 7 & 10 \\
            \hline
        \end{tabular}
        \caption*{GAEN None Group Trust}
    }
    \vspace{-0.7cm}
\end{table}

When we break their trust scores up by release organization, we see that of those who were told Google-Apple developed the application, 18 would trust the application, 25 were unsure, and 8 would not. Compared with 18 who would trust the application more if a government organization released it, 10 who were unsure, and 23 who would not trust a government-released application. Only 8 of our Google-Apple group respondents would both trust Google-Apple and trust a government application more. 

\begin{table}[ht]
    \parbox{.30\linewidth}{
        \centering
        \begin{tabular}{|c|c|c|c|}
            \hline
            & Yes & Unsure & No \\
            \hline
            GAEN & 14 & 19 & 17 \\
            \hline
            Gov & 16 & 27 & 7 \\
            \hline
        \end{tabular}
        \caption*{Gov Groups Trust}
    }
    \hfill
    \parbox{.30\linewidth}{
        \centering
        \begin{tabular}{|c|c|c|c|}
            \hline
            & Yes & Unsure & No \\
            \hline
            GAEN & 5 & 9 & 9 \\
            \hline
            Gov & 6 & 13 & 4 \\
            \hline
        \end{tabular}
        \caption*{Gov Explanation Group Trust}
    }
    \hfill
    \parbox{.30\linewidth}{
        \centering
        \begin{tabular}{|c|c|c|c|}
            \hline
            & Yes & Unsure & No \\
            \hline
            GAEN & 9 & 10 & 8 \\
            \hline
            Gov & 10 & 14 & 3 \\
            \hline
        \end{tabular}
        \caption*{Gov None Group Trust}
    }
    \vspace{-0.7cm}
\end{table}

Of the group who received the government statement, 16 said they would trust it, 27 were unsure, and 7 would not trust the application. This is compared to 14 who said they would trust an application released by Google-Apple more, 19 who were unsure, and 17 who would not trust the alternative. Here we see 7 participants who would trust the government application but would trust a private application more.

\begin{table}[ht]
    \centering
    \begin{tabular}{|c|c|c|c|c|}
        \hline
         & No & Unsure & Yes & T-test \\
        \hline
        Gov (Exp) & 9 & 9 & 5 & \multirow{2}{*}{0.03407}\\
        GAEN (Exp) & 3 & 10 & 10 & \\
        \hline
        \textbf{Total} & 12 & 19 & 15 & \cellcolor[rgb]{0,0,0} " " \\
        \hline
    \end{tabular}
    \hfill
    \centering
    \begin{tabular}{|c|c|c|c|c|}
        \hline
         & No & Unsure & Yes & T-test \\
        \hline
        Gov (None) & 8 & 10 & 9 & \multirow{2}{*}{0.7304}\\
        GAEN (None) & 5 & 15 & 8 & \\
        \hline
        \textbf{Total} & 13 & 25 & 17 & \cellcolor[rgb]{0,0,0} " " \\
        \hline
    \end{tabular}
    \caption{GAEN Application Trust}
    \vspace{-0.8cm}
\end{table}

We then examined the GAEN application after receiving the explanation. For this, we compared the trust from the government group of the alternative application with the trust levels of the GAEN group with their given application. When we run a t-test on these statistics, we see a p-value of 0.03407 (t = 2.1881, df = 43.558) and 0.03794 from a Wilcoxon test. When instead looking at the participants who did not receive the explanation, we see a p-value of 0.7304 (t = 0.34651, df = 50.971) from a t-test and 0.765 from a Wilcoxon test. 

\begin{table}[ht]
    \centering
    \begin{tabular}{|c|c|c|c|c|}
        \hline
         & No & Unsure & Yes & T-test \\
        \hline
        Gov (Exp) & 4 & 13 & 6 & \multirow{2}{*}{0.1489}\\
        GAEN (Exp) & 13 & 3 & 7 & \\
        \hline
        \textbf{Total} & 12 & 19 & 15 & \cellcolor[rgb]{0,0,0} " " \\
        \hline
    \end{tabular}
    \hfill
    \centering
    \begin{tabular}{|c|c|c|c|c|}
        \hline
         & No & Unsure & Yes & T-test \\
        \hline
        Gov (None) & 3 & 14 & 10 & \multirow{2}{*}{0.2899}\\
        GAEN (None) & 10 & 7 & 11 & \\
        \hline
        \textbf{Total} & 13 & 25 & 17 & \cellcolor[rgb]{0,0,0} " " \\
        \hline
    \end{tabular}
    \caption{Government Application Trust}
    \vspace{-0.8cm}
\end{table}

Looking at the overall trust in the government application, we assessed the trust of the GAEN group for the alternative application compared to the government group’s trust in their given application. For those who received the explanation, we see a p-value of 0.1489 (t = 1.4718, df = 40.264) from a t-test and a p-value of 0.1205 from a Wilcoxon ranked sum test. When examining those who did not receive the explanation, a t-test returned a p-value of 0.2899 (t = 1.0698, df = 49.85), while the Wilcoxon test returned a p-value of 0.3636. 

Next, we compared the trust in an application dependent on the releasing organization. When we look at the trust of the given application, we see a p-value of 0.9053 (t = 0.11933, df = 98.92) from a t-test, a p-value of 0.875 from a Wilcoxon test, and a p-value of 0.8819 (X-squared = 0.25136, df = 2) from a chi-squared test. When looking at the trust of the alternative application, we see a p-value of 0.8221 (t = 0.22546, df = 97.899) from the t-test, a p-value of 0.7724 from a Wilcoxon test, and a p-value of 0.1235 (X-squared = 4.1836, df = 2) from a chi-squared test.

We then broke this down further, looking only at those that received the explanation. Beginning with their given organization, a t-test comparing the GAEN and Government group's trust after receiving the explanation returned a p-value of 0.2883 (t = 1.0749, df = 43.888), and a Wilcoxon test returned 0.2673. A chi-square test returned a warning that the p-value may not be reliable but returned 0.4644 (X-squared = 1.5342, df = 2). When we turn to the alternative application for the same subdivision, we see a p-value of 0.7301 (t = -0.3472, df = 42.88) being returned by the t-test, a p-value of 0.6096 being returned by the Wilcoxon test, and a p-value of 0.1313 (X-squared = 4.0606, df = 2) being returned from a chi-squared test. 

We see similar results when we examine those who did not receive the explanation. For their given application, a t-test returned a p-value of 0.404 (t = 0.84116, df = 52.998), while a Wilcoxon ranked sum test gave 0.4121. Again, the chi-squared test returned a warning but gave the p-value of 0.6912 (X-squared = 0.73877, df = 2). For the alternative application, we see a p-value of 0.9954 (t = 0.0058066, df = 52.869) from a t-test, a p-value of 1 from a Wilcoxon test, and a p-value of 0.627 (X-squared = 0.93376, df = 2) from a chi-squared test.

\subsection{Does Privacy-Consciousness Effect Trust}
\begin{wraptable}{l}{6cm}
    \centering
    \begin{tabular}{|c|c|c|}
        \hline
        Privacy Scale & Total & Cumulative\\
        \hline
        <50\% & 3 & \multirow{2}{*}{14}\\
        50-60\% & 11 & \\
        \hline
        60-70\% & 21 & 35 \\\cline{2-3}
        \hline
        70-80\% & 37 & 72 \\\cline{2-3}
        \hline
        80-90\% & 19 & 91 \\\cline{2-3}
        \hline
        90-100\% & 4 & 95 \\\cline{2-3}
        \hline
    \end{tabular}
    \caption{Privacy-Conscious Scores}
    \vspace{0.1cm}
    \centering
    \begin{tabular}{|c|c|}
        \hline
        Classification & Total \\
        \hline
        Private & 78 \\
        \hline
        Open & 21 \\
        \hline
    \end{tabular}
    \caption{Self-Described Privacy}
    \vspace{-0.8cm}
\end{wraptable}

Looking at the scale laid out by Egelman, we see that most of our participants scored well on the metrics. We had 6 respondents who did not answer all sections of the scale, so these have been discounted for this section of the analysis, leaving us with 95 respondents. All questions were scored on a 5-point Likert scale, with ‘definitely yes’ earning a 5 and ‘definitely no’ earning a 1 for the questions where yes was the more privacy-conscious option. The inverse was true for those questions where no was the more appropriate answer. 92 of our 95 eligible participants scored over 50\% (97\% of the population), 60 scored over 70\% (63\% of the population) and 4 scored over 90\% (4\% of the population). 

We also see that 78 of our 99 respondents (2 participants elected not to answer) responded that they view themselves as “a more private person and like to keep to myself”. In contrast, 21 viewed themselves as “a more open person who enjoys sharing with others”.  

We do see interesting results when we separate respondents by their self-description of their privacy habits, whether they are a “private” person or an “open” person. Ignoring release organization or whether participants received the explanation, we see a p-value of 0.0107 (t = -2.6996, df = 34.318) from a t-test, 0.01161 from a Wilcoxon test and 0.06391 (X-squared = 5.5005, df = 2) from a chi-squared test, although this did return the same warning as the earlier samples. Significant p-values are not seen when separating by release organization or whether the respondent received the explanation. 

Our final comparison was between the privacy scores of participants and their willingness to trust an application. We classified those who scored less than or equal to 70\% as having ‘poor’ privacy habits and those who scored over 70\% as having ‘good’ privacy habits. When we consider their given application, we see a p-value of 0.3838 (t = 0.87532, df = 85.693) from a t-test and a p-value of 0.5281 from a Wilcoxon test. When we run a chi-squared test, we see a p-value of 0.1143 (X-squared = 4.3374, df = 2). When we instead look at the alternative application, we see a p-value of 0.7402 (t = 0.33281, df = 71.775) returned from a t-test, a p-value of 0.7332 returned from a Wilcoxon test, and a p-value of 0.905 (X-squared = 0.19973, df = 2) returned from a chi-squared test. 

\section{Discussion}
In this section, we will take a deeper look into the conclusions that can be drawn from the results presented in the above section.

\subsection{Limitations}
There were a few limitations in our study, which may have led to some of the inclusive results drawn from our participants. Given our survey only sought 100 respondents, we do not have a true representation of the US population. To obtain an accurate sample with a confidence level of 95\% and an error margin of 5\%, we would have had to survey 4 times the participants we were able to. The decision to run this small of a sample size was due to budgeting restrictions. We decided not to increase our result size with further rounds of participants after analysis of the current data suggested a lack of correlation between receiving the explanation and increased trust, which we did not perceive to be different with a larger sample size. Suppose we lowered our error margin to 10\%. In that case, we see that our sample size would be acceptable for the US population, another determining factor in our decision not to sample a larger population. Given these limitations, it is possible that a larger sample size may produce different results. Still, given the large p-values returned from our various testing functions, we do not believe this to be likely.

\subsection{Situational Trust Scenarios}
Of our three hypothetical scenarios presented to participants, celebrity endorsement was the scenario with the most negligible effect. Here we see only 14 respondents said they probably or definitely would download the application if a celebrity endorsed it. All 14 of these participants would also probably or definitely download the application voluntarily, and if required by their local government, suggesting they are more trusting of the application. They also all responded ‘unsure’ or ‘yes’ to trusting their given application. Although is unclear if those participants who answered probably or definitely not to this question would reconsider their decision to download based on which celebrity was endorsing it or if they believe a celebrity endorsement would not affect their willingness to download. 

The scenario with the highest download rate was when required by a local government. Here we see an even split between release organizations. When we delve into this group more closely, we see that most respondents were either unsure or would trust their given application. However, two of those who would download the application if required by their local government would not trust the application. These two respondents were from different groups, with one receiving the government explanation and the other not receiving an explanation as a part of the GAEN group. Both would not trust their given application or the alternative and had no similar demographics. From the responses to this question, we can conclude that government mandating may positively affect download rates but are unlikely to increase trust. 

\subsection{RQ1: Does Understanding Affect User Trust}
Firstly, looking at the comprehension accuracy of those who received the explanation, we see 17 of our 46 participants scored 70\% or over on the comprehension questions. However, we do not see a relationship when we examine whether those who understood the explanation better were more likely to trust the application. This strongly suggests we cannot assume participant understanding had a direct impact on their willingness to trust the application. It was also clear from the results in our pilot tests that it was likely that users carried their own bias through the survey, regardless of what was contained in the explanation. .  

When we look at the results from the tallies, we see a fair amount of uncertainty from respondents regarding their willingness to trust a contact tracing application. In 3 out of the 4 groups, we see 'unsure' as the most popular response for the trust of their given application, with the only exception being the GAEN explanation group, who tied ‘unsure’ with ‘yes’. This lack of certainty from our participants did not seem to be affected by the explanation. When we delve deeper into those who received the explanation but were still unsure, we see a range of comprehension scores, meaning that we cannot correlate the understanding of this subset with their uncertainty regarding trusting a contact tracing application. 

\begin{wraptable}{l}{5cm}
    \begin{tabular}{|c|c|c|c|}
    \hline
         & Total & Explain & None \\
         \hline
        GAEN & 32 & 15 & 17 \\
        \hline
        Gov & 34 & 13 & 21 \\
        \hline
    \end{tabular}
    \caption*{Preferred Application}
    \vspace{-0.7cm}
\end{wraptable}

If we remove those who responded as uncertain, we still cannot see a strong correlation between explanation and trust. In all four cases, we notice a higher percentage of those who would trust their given application than the alternative, regardless of whether they received the explanation. This statistic has little bearing, as when we examine only those who answered ‘yes’ to trusting either their given application or the alternative application, we do not see strong evidence for a favored release organization. When t-tests were run on various scenarios which removed those answering ‘unsure’, we still saw no p-values that definitively rejected the null hypothesis.

Once we look at the effects of the explanation on willingness to trust, regardless of how the data is separated, we do not see strong evidence to support our first hypothesis. Separating by whether participants received the explanation yields a relatively high p-value (>0.9) for their given application. While the p-value for the alternate application was closer to supporting our hypothesis, it still was not within the acceptable range ($\sim$0.1). We could theorize that the explanation caused participants to prefer the alternative application instead; perhaps because they only understand the way their given application works, they see the alternative as a better option of the two. However, this conclusion does not hold when we separate the responses by production organization and then examine the effect of the explanation. For both the government and GAEN applications, we see similar p-values for both the given application and the alternative (between 0.2 and 0.3), suggesting the theory presented above does not hold. 

\subsection{RQ2: Does Release Organization Affect User Trust?}
If we instead look at the effect of the release organization on willingness to trust, we do not see support for our second hypothesis. Separating by release organization alone, we see a high p-value (>0.8) for both their given and alternative applications. This provides strong evidence that we cannot confirm our hypothesis, as there is no strong correlation between release organization and trust. This observation is further confirmed when we examine both release organizations separated by whether the participants received the explanation. Of the four cases, the closest to a p-value of less than 0.05 was those who received the explanation’s trust of their given application, but this still returned a p-value greater than 0.2 and, as a result, does not provide strong evidence of a correlation.

If we examine by application, we also cannot find a significant correlation between the release organization, understanding, and user trust. Of the four scenarios, only the GAEN application presented us with statistically significant p-values. Still, given the lack of significance for other scenarios, it is unlikely this information holds much relevance. It is possible, given the framing of the question, the situations where participants were asked about their preferences for the alternative application may not return accurate results. We assumed that if a participant responded no to the alternative application, they would not trust the application. It is possible they responded no because they trusted their given application more, and these responses were discounted unfairly. This possibility would require further research to confirm.  

\subsection{Does Privacy-Consciousness Effect Trust}
We are presented with questionable relevance when we investigate the effect of self-privacy rating on their willingness to trust the applications. 78 of our respondents self-reported as being a ‘private’ person, and 21 self-reported as an ‘open’ person. When we separate by this classification, we see a strong correlation between a person’s self-classification and willingness to trust their given application (p-value = 0.0107, t = -2.6996, df = 34.318). We also see similar numbers for the alternative application. However, when we further separate these groups by either release organization or whether they received the explanation, we do not see similar results, suggesting this statistical significance does not bear any actual weight. 

Regarding our privacy scale, 60 of our 95 qualifying respondents scored over 70\% and were classified as having ‘good’ privacy. However, no analysis returned a significant p-value (all were over 0.3). As a result, we cannot conclude that those with good privacy standards are more or less likely to trust a contact tracing application.

Overall, neither of our hypotheses were supported by the data collected. Our first research question, investigating whether understanding the technology behind the Google-Apple Exposure Notification API affects users’ trust in such an application, drew inconclusive results. While this does not disprove our hypothesis, it does not definitively confirm it either. Our second research question, investigating the impact of private companies in these applications on users’ trust, drew similar inconclusive answers. Again, although the results do not disprove our hypothesis, nor can we prove our hypothesis definitively. This would require further work and a larger sample size to investigate these research questions further.

\subsection{Future Research}
An area for further research could be to examine the reasons why those who downloaded their state’s COVID-19 application no longer have it, and whether this was affected by the application itself or the current state of the pandemic. The prevalence of the COVID-19 pandemic has also taken a back-seat in the minds of the general public, and this study could have different findings were it performed a year or two earlier, during the height of 2020 lockdowns.

\section{Conclusion}
In this paper, we present statistical research into both the effects of both the release organization and the public’s understanding of privacy protection architecture in COVID-19 contact tracing applications on their willingness to trust such applications. 101 survey responses were collected in June of 2022 from respondents located in the United States. We found that our results were unable to definitively answer either research question, as there were no significant indicators of correlation between either release organization and trust or understanding and trust.


\bibliographystyle{ACM-Reference-Format}
\bibliography{main}


\begin{thebibliography}{13}


\ifx \showCODEN    \undefined \def \showCODEN     #1{\unskip}     \fi
\ifx \showDOI      \undefined \def \showDOI       #1{#1}\fi
\ifx \showISBNx    \undefined \def \showISBNx     #1{\unskip}     \fi
\ifx \showISBNxiii \undefined \def \showISBNxiii  #1{\unskip}     \fi
\ifx \showISSN     \undefined \def \showISSN      #1{\unskip}     \fi
\ifx \showLCCN     \undefined \def \showLCCN      #1{\unskip}     \fi
\ifx \shownote     \undefined \def \shownote      #1{#1}          \fi
\ifx \showarticletitle \undefined \def \showarticletitle #1{#1}   \fi
\ifx \showURL      \undefined \def \showURL       {\relax}        \fi
\providecommand\bibfield[2]{#2}
\providecommand\bibinfo[2]{#2}
\providecommand\natexlab[1]{#1}
\providecommand\showeprint[2][]{arXiv:#2}

\bibitem[Cry(2020)]%
        {Cryptography}
 \bibinfo{year}{2020}\natexlab{}.
\newblock \bibinfo{title}{Privacy-preserving contact tracing - apple and
  google}.
\newblock
\newblock
\urldef\tempurl%
\url{https://covid19.apple.com/contacttracing}
\showURL{%
\tempurl}


\bibitem[Altmann et~al\mbox{.}(2020)]%
        {Altmann}
\bibfield{author}{\bibinfo{person}{Samuel Altmann}, \bibinfo{person}{Luke
  Milsom}, \bibinfo{person}{Hannah Zillessen}, \bibinfo{person}{Raffaele
  Blasone}, \bibinfo{person}{Frederic Gerdon}, \bibinfo{person}{Ruben Bach},
  \bibinfo{person}{Frauke Kreuter}, \bibinfo{person}{Daniele Nosenzo},
  \bibinfo{person}{S{\'e}verine Toussaert}, {and} \bibinfo{person}{Johannes
  Abeler}.} \bibinfo{year}{2020}\natexlab{}.
\newblock \showarticletitle{Acceptability of app-based contact tracing for
  COVID-19: Cross-country survey study}.
\newblock \bibinfo{journal}{\emph{JMIR mHealth and uHealth}}
  \bibinfo{volume}{8}, \bibinfo{number}{8} (\bibinfo{year}{2020}),
  \bibinfo{pages}{e19857}.
\newblock


\bibitem[Calloway et~al\mbox{.}(2020)]%
        {Calloway}
\bibfield{author}{\bibinfo{person}{Laura Calloway}, \bibinfo{person}{Hilda
  Hadan}, \bibinfo{person}{Shakthidhar Gopavaram}, \bibinfo{person}{Shrirang
  Mare}, {and} \bibinfo{person}{L~Jean Camp}.} \bibinfo{year}{2020}\natexlab{}.
\newblock \showarticletitle{Privacy in Crisis: Participants' Privacy
  Preferences for Health and Marketing Data during a Pandemic}. In
  \bibinfo{booktitle}{\emph{Proceedings of the 19th Workshop on Privacy in the
  Electronic Society}}. \bibinfo{pages}{181--189}.
\newblock


\bibitem[Chan and Saqib(2021)]%
        {ChanSaqib}
\bibfield{author}{\bibinfo{person}{Eugene~Y. Chan} {and}
  \bibinfo{person}{Najam~U. Saqib}.} \bibinfo{year}{2021}\natexlab{}.
\newblock \showarticletitle{Privacy concerns can explain unwillingness to
  download and use contact tracing apps when COVID-19 concerns are high}.
\newblock \bibinfo{journal}{\emph{Computers in Human Behavior}}
  \bibinfo{volume}{119} (\bibinfo{year}{2021}), \bibinfo{pages}{106718}.
\newblock
\showISSN{0747-5632}
\urldef\tempurl%
\url{https://doi.org/10.1016/j.chb.2021.106718}
\showDOI{\tempurl}


\bibitem[Egelman and Peer(2015)]%
        {Egelman}
\bibfield{author}{\bibinfo{person}{Serge Egelman} {and} \bibinfo{person}{Eyal
  Peer}.} \bibinfo{year}{2015}\natexlab{}.
\newblock \showarticletitle{Scaling the security wall: Developing a security
  behavior intentions scale (sebis)}. In \bibinfo{booktitle}{\emph{Proceedings
  of the 33rd annual ACM conference on human factors in computing systems}}.
  \bibinfo{pages}{2873--2882}.
\newblock


\bibitem[Grande et~al\mbox{.}(2021)]%
        {Grande}
\bibfield{author}{\bibinfo{person}{David Grande}, \bibinfo{person}{Nandita
  Mitra}, \bibinfo{person}{Xochitl~Luna Marti}, \bibinfo{person}{Raina
  Merchant}, \bibinfo{person}{David Asch}, \bibinfo{person}{Abby Dolan},
  \bibinfo{person}{Meghana Sharma}, {and} \bibinfo{person}{Carolyn Cannuscio}.}
  \bibinfo{year}{2021}\natexlab{}.
\newblock \showarticletitle{Consumer Views on Using Digital Data for COVID-19
  Control in the United States}.
\newblock \bibinfo{journal}{\emph{JAMA network open}} \bibinfo{volume}{4},
  \bibinfo{number}{5} (\bibinfo{year}{2021}),
  \bibinfo{pages}{e2110918--e2110918}.
\newblock


\bibitem[Kaptchuk et~al\mbox{.}(2020)]%
        {Kaptchuk}
\bibfield{author}{\bibinfo{person}{Gabriel Kaptchuk},
  \bibinfo{person}{Daniel~G. Goldstein}, \bibinfo{person}{Eszter Hargittai},
  \bibinfo{person}{Jake Hofman}, {and} \bibinfo{person}{Elissa~M. Redmiles}.}
  \bibinfo{year}{2020}\natexlab{}.
\newblock \bibinfo{title}{How good is good enough for COVID19 apps? The
  influence of benefits, accuracy, and privacy on willingness to adopt}.
\newblock
\newblock
\showeprint[arxiv]{2005.04343}~[cs.CY]


\bibitem[Li et~al\mbox{.}(2021)]%
        {Li}
\bibfield{author}{\bibinfo{person}{Tianshi Li}, \bibinfo{person}{Camille Cobb},
  \bibinfo{person}{Jackie}, \bibinfo{person}{Yang}, \bibinfo{person}{Sagar
  Baviskar}, \bibinfo{person}{Yuvraj Agarwal}, \bibinfo{person}{Beibei Li},
  \bibinfo{person}{Lujo Bauer}, {and} \bibinfo{person}{Jason~I. Hong}.}
  \bibinfo{year}{2021}\natexlab{}.
\newblock \bibinfo{title}{What Makes People Install a COVID-19 Contact-Tracing
  App? Understanding the Influence of App Design and Individual Difference on
  Contact-Tracing App Adoption Intention}.
\newblock
\newblock
\showeprint[arxiv]{2012.12415}~[cs.HC]


\bibitem[Maytin et~al\mbox{.}(2021)]%
        {Maytin}
\bibfield{author}{\bibinfo{person}{Lauren Maytin}, \bibinfo{person}{Jason
  Maytin}, \bibinfo{person}{Priya Agarwal}, \bibinfo{person}{Anna Krenitsky},
  \bibinfo{person}{JoAnn Krenitsky}, {and} \bibinfo{person}{Robert~S Epstein}.}
  \bibinfo{year}{2021}\natexlab{}.
\newblock \showarticletitle{Attitudes and perceptions toward COVID-19 digital
  surveillance: survey of young adults in the United States}.
\newblock \bibinfo{journal}{\emph{JMIR formative research}}
  \bibinfo{volume}{5}, \bibinfo{number}{1} (\bibinfo{year}{2021}),
  \bibinfo{pages}{e23000}.
\newblock


\bibitem[Seberger and Patil(2021)]%
        {Seberger}
\bibfield{author}{\bibinfo{person}{John~S. Seberger} {and}
  \bibinfo{person}{Sameer Patil}.} \bibinfo{year}{2021}\natexlab{}.
\newblock \showarticletitle{Us and Them (and It): Social Orientation, Privacy
  Concerns, and Expected Use of Pandemic-Tracking Apps in the United States}.
  In \bibinfo{booktitle}{\emph{Proceedings of the 2021 CHI Conference on Human
  Factors in Computing Systems}} (Yokohama, Japan) \emph{(\bibinfo{series}{CHI
  '21})}. \bibinfo{publisher}{Association for Computing Machinery},
  \bibinfo{address}{New York, NY, USA}, Article \bibinfo{articleno}{65},
  \bibinfo{numpages}{19}~pages.
\newblock
\showISBNx{9781450380966}
\urldef\tempurl%
\url{https://doi.org/10.1145/3411764.3445485}
\showDOI{\tempurl}


\bibitem[Triandis(2001)]%
        {Triandis}
\bibfield{author}{\bibinfo{person}{Harry~C. Triandis}.}
  \bibinfo{year}{2001}\natexlab{}.
\newblock \showarticletitle{Individualism-Collectivism and Personality}.
\newblock \bibinfo{journal}{\emph{Journal of Personality}}
  \bibinfo{volume}{69}, \bibinfo{number}{6} (\bibinfo{year}{2001}),
  \bibinfo{pages}{907--924}.
\newblock
\urldef\tempurl%
\url{https://doi.org/10.1111/1467-6494.696169}
\showDOI{\tempurl}
\showeprint{https://onlinelibrary.wiley.com/doi/pdf/10.1111/1467-6494.696169}


\bibitem[Wang et~al\mbox{.}(2021)]%
        {WangBashir}
\bibfield{author}{\bibinfo{person}{Tian Wang}, \bibinfo{person}{Lin Guo}, {and}
  \bibinfo{person}{Masooda Bashir}.} \bibinfo{year}{2021}\natexlab{}.
\newblock \showarticletitle{COVID-19 Apps and Privacy Protections from Users'
  Perspective}.
\newblock \bibinfo{journal}{\emph{Proceedings of the Association for
  Information Science and Technology}} \bibinfo{volume}{58},
  \bibinfo{number}{1} (\bibinfo{year}{2021}), \bibinfo{pages}{357--365}.
\newblock
\urldef\tempurl%
\url{https://doi.org/10.1002/pra2.463}
\showDOI{\tempurl}
\showeprint{https://asistdl.onlinelibrary.wiley.com/doi/pdf/10.1002/pra2.463}


\bibitem[Zhang et~al\mbox{.}(2020)]%
        {Zhang}
\bibfield{author}{\bibinfo{person}{Baobao Zhang}, \bibinfo{person}{Sarah
  Kreps}, \bibinfo{person}{Nina McMurry}, {and} \bibinfo{person}{R~Miles
  McCain}.} \bibinfo{year}{2020}\natexlab{}.
\newblock \showarticletitle{Americans’ perceptions of privacy and
  surveillance in the COVID-19 pandemic}.
\newblock \bibinfo{journal}{\emph{Plos one}} \bibinfo{volume}{15},
  \bibinfo{number}{12} (\bibinfo{year}{2020}), \bibinfo{pages}{e0242652}.
\newblock


\end{thebibliography}

\appendix
\newpage
\section{Final Explanation}
The Google-Apple Exposure Notification architecture was created to allow for cross-platform detection of COVID-19 exposure. This technology was created with privacy as a foremost concern.
    
In order to maintain privacy, this architecture DOES NOT utilize GPS tracking, using Bluetooth technology to detect nearby devices, but never recording location.

When a device signs up, it is allocated a unique identifier. The device also has a temporary key, which changes at regular intervals to protect privacy, and is based on the original identifier to make it able to be recognized later.

When two devices come within Bluetooth range of each other, they ask each other for their current temporary key. They also store information about the interaction like the duration the device was in range and the strength of the connection. This is so if a positive test occurs, the device can determine the risk to the individual.

If a person tests positive and reports in the app, their device uploads their unique identifier, as well as the intervals at which they may have been contagious. When other devices poll the central server they are able to determine, based on the identifier and intervals, if any of the temporary keys they have saved refer to a positive case.

If a person never tests positive or chooses not to report, this information remains only on their own device, making the temporary keys on other devices undecipherable.

\textbf{The following example explains the basics of how the system functions in the real world. In a real application, the identifiers and keys are more complex to promote privacy, this example is simplified for ease of understanding.}

Anna’s Unique Identifier: 1234\\
Bob’s Unique Identifier: 5678

Interval Logic: Every day the temporary key adds 1 to the value stored, beginning with the unique identifier.

\begin{itemize}
    \item Bob and Anna cross paths in a coffee shop
    \item Bob’s phone saves 1244 as a key it came in contact with
    \item Anna’s phone saves 5687 as a key it came in contact with
    \item Anna tests positive for COVID and chooses to report
    \item Anna’s phone uploads 1234, and the numbers [9-14] for the days in which she was contagious.
    \item Bob’s phone does its regular poll of the central server. It sees that 1234+10=1244, which is a key it has stored
    \item Bob’s phone reviews its data from that encounter and determines that Bob was at risk
    \item Bob receives a notification that he may have been exposed to COVID-19 and should isolate, but receives no information about Anna, as Bob’s phone does not know any personal information about Anna. 
\end{itemize}

\end{document}